\documentclass[12pt]{elsart}
\usepackage{epsfig,graphics}

\newlength{\lslash}
\def\slash#1{\settowidth{\lslash}{$#1$}\makebox[\lslash]{\makebox[0mm]{$/$}\makebox[0mm]{$#1$}}}

\newcommand{\pslash}{\FMslash p}

\newcommand{\chiral}[1]{\stackrel{\circ}{#1}}

\begin{document}
\begin{frontmatter}
\title{On the possibility \\of a discontinuous quark-mass dependence \\ of baryon octet and decuplet masses}
\author[GSI]{A. Semke}
\author[GSI]{and M.F.M. Lutz}
\address[GSI]{Gesellschaft f\"ur Schwerionenforschung (GSI)\\
Planck Str. 1, 64291 Darmstadt, Germany}
\begin{abstract}
We compute the quark-mass dependence of the baryon octet and decuplet masses within the
$\chi$-MS scheme at the one-loop level. The results are confronted with recent lattice QCD
simulations of the MILC collaboration. A fair reproduction of the quark-mass dependence
as suggested by the lattice simulations is obtained up to pion masses of about $700$ MeV.
Based on the chiral one-loop results we predict that the  dependence of the baryon octet
and decuplet masses on the quark-masses is discontinuous. Typically the pion-mass dependence
is smooth up to about 300 MeV only. This is a consequence of
self consistency imposed on the partial summation, i.e. the masses used in the loop functions
are identical to those obtained from the baryon self energies.
The 'mysterious' quark-mass dependence of the $\Xi$ mass predicted by the MILC
collaboration is recovered in terms of a discontinuous chiral extrapolation.
\end{abstract}
\end{frontmatter}

\section{Introduction}

The present-day interpretation of QCD lattice simulations requires a profound understanding of
the dependence of observable quantities on the light quark masses.
A powerful tool to derive such dependencies
is the chiral Lagrangian, an effective field theory based on the chiral properties of QCD.
The application of strict chiral perturbation theory to the SU(3) flavor sector of QCD is plagued
by poor convergence properties for processes involving
baryons \cite{Jenkins:1992,Bernard:Kaiser:Meissner:1993,Borasoy:Meissner:1997,Ellis:Torikoshi:1999,Lehnhart:Gegelia:Scherer:2005}.
Thus it is important to establish partial summation schemes that enjoy improved convergence properties and that are
better suited for chiral extrapolations of lattice simulations.
It was suggested by Donoghue and Holstein
\cite{Donoghue:Holstein:1998,Donoghue:Holstein:Borasoy:1999,Borasoy:Holstein:Lewis:Ouimet:2002}
to improve the convergence properties by introducing a finite cutoff into the heavy-baryon effective field theory of
Jenkins and Manohar \cite{Jenkins:Manohar:1991}.
Another scheme was suggested in \cite{Lutz:2000,Lutz:Kolomeitsev:2002,Semke:Lutz:2005}, the construction
of which was guided by keeping manifestly covariance, analyticity and causality. The computation of the baryon
octet and decuplet masses in this scheme was worked out recently indicating convincing convergence
properties of the chiral loop expansion \cite{Semke:Lutz:2005}.

For studies of chiral extrapolations of the nucleon mass within
the chiral SU(2) Lagrangian we refer to the work by Procura,
Hemmert and Weise \cite{Muenchen:2004}. The latter study applied
the partial summation scheme of Becher and Leutwyler
\cite{Becher:Leutwyler:1999}. It  was based on simulations of the
CP-PACS collaboration that used dynamical u- and d-quarks only
\cite{CP-PACS:2003}. The application of the scheme by Becher and
Leutwyler \cite{Becher:Leutwyler:1999} to the chiral SU(3)
Lagrangian is questioned by poor convergence properties as was
demonstrated by Ellis and Torikoshi at hand of the baryon octet
masses \cite{Ellis:Torikoshi:1999}. The recent work of Pascalutsa
and Vanderhagen \cite{Pascalutsa:Vanderhagen:2005} suggested to
take seriously the partial summation implied by the  extended
on-mass shell scheme (EOMS) introduced by Gegelia and Japaridze
\cite{Gegelia:Japaridze:1999}. Their computation addressed so far
the chiral extrapolation of the nucleon and isobar mass only.

It is the purpose of the present work to confront the results of \cite{Semke:Lutz:2005} with recent lattice QCD
simulation of the MILC collaboration \cite{MILC:2001,MILC:2004}, that use dynamical u-,d- and s-quarks in the
staggered approximation. We do not aim at a quantitative extrapolation
of the lattice simulation, which would require a continuum limit extrapolation and a quantitative control of
systematic errors.  Rather, we want to use the available partial lattice results to make predictions
for future QCD lattice simulations and possibly obtain rough constraints on some chiral parameters.
An analysis of the recent MILC results, similar in spirit, was undertaken by Frink and
Mei\ss ner \cite{Frink:Meissner:2005} based on the cutoff scheme of Donoghue and Holstein.

Adjusting the values of the $Q^2$ counter terms to the physical masses we predict a discontinuous
dependence of the baryon masses on the pion mass. This is a consequence of self consistency imposed on the
partial summation approach, i.e.
the masses used in the loop function are identical to those obtained from the baryon self energy.
The latter is a crucial requirement since the loop functions depend sensitively on the
precise values of the baryon masses. Our results may explain the mysterious quark-mass dependence
observed by the MILC collaboration for the $\Xi$ mass \cite{MILC:2001,MILC:2004}.

\section{Chiral interaction terms}

We collect the terms of the chiral Lagrangian that determine the leading orders of
baryon octet and decuplet self energies \cite{Krause:1990,Bernard:Hemmert:Meissner:2003}.
Up to chiral order $Q^2$ the baryon propagators follow from
\begin{eqnarray}
{\mathcal L} &=&  \mathrm{tr} \,\Big( \bar{B}\,\big[i\, \slash{\partial}\,-
\stackrel{\circ}{M}_{[8]}\big]\,B \Big)
\nonumber \\
&-& \mathrm{tr}\,\Big(\bar{\Delta}_\mu \cdot \Big(\big[i\,\slash{\partial}\,
-\stackrel{\circ}{M}_{[10]}\big]\,g^{\mu\nu} -i\,(\gamma^\mu \partial^\nu + \gamma^\nu \partial^\mu)
+ \gamma^\mu\,\big[i\,\slash{\partial} + \stackrel{\circ}{M}_{[10]}\big]\,\gamma^\nu \Big)\,\Delta_\nu\Big)
\nonumber \\
& -& 2\,d_0\, \mathrm{tr}\Big(\bar{\Delta}_\mu \cdot \Delta^\mu \Big)\, \mathrm{tr}\Big(\chi_0\Big)
-2\,d_D\, \mathrm{tr}\Big( (\bar{\Delta}_\mu \cdot \Delta^\mu)\, \chi_0\Big)
\nonumber \\
&+&2\,b_0 \,\mathrm{tr}\Big(\bar{B}\,B\Big)\, \mathrm{tr}\Big(\chi_0\Big)
+ 2\,b_F\,\mathrm{tr}\Big(\bar{B}\,[\chi_0,B]\Big) +
2\,b_D\,\mathrm{tr}\Big(\bar{B}\,\{\chi_0,B\}\Big) \,,
\nonumber\\
\nonumber\\
&& \chi_0 = \left( \begin{array}{ccc}
m_\pi^2 & 0 & 0 \\
0 & m_\pi^2 & 0 \\
0 & 0 & 2\,m_K^2-m_\pi^2
\end{array}\right)\,.
\label{chiral-L}
\end{eqnarray}
We  assume perfect isospin symmetry through out
this work. The  fields are decomposed
into isospin multiplets
\begin{eqnarray}
\Phi &=& \tau \cdot  \pi + \alpha^\dagger \!\cdot \! K + K^\dagger
\cdot \alpha  + \eta \,\lambda_8 \;,
\nonumber\\
\sqrt{2}\,B &=&  \alpha^\dagger \!\cdot \! N+\lambda_8 \,\Lambda+ \tau \cdot \Sigma
 +\Xi^t\,i\,\sigma_2 \!\cdot \!\alpha   \, ,
\end{eqnarray}
with the  Gell-Mann matrices, $\lambda_i$, and the isospin doublet fields
$K =(K^+,K^0)^t $ and $\Xi = (\Xi^0,\Xi^-)^t$.
The isospin Pauli matrices $\sigma=(\sigma_1,\sigma_2,\sigma_3)$ act
exclusively in the space of isospin doublet fields $(K,N,\Xi)$ and
the matrix valued isospin doublet $\alpha$,
\begin{eqnarray}
&& \alpha^\dagger =
 {\textstyle{1\over\sqrt{2}}}\left( \lambda_4+i\,\lambda_5 ,
\lambda_6+i\,\lambda_7 \right) \;,\;\;\;\tau =
(\lambda_1,\lambda_2,\lambda_3)\;.
 \label{def-alpha}
\end{eqnarray}

The evaluation of the baryon self energies to order $Q^3$ probes the meson-baryon vertices
\begin{eqnarray}
{\mathcal L} &=& \frac{F}{2f}\, \mathrm{tr} \,\Big( \bar{B}\, \gamma_5 \gamma^\mu \,[\partial_\mu \Phi,\,B] \Big)
+ \frac{D}{2f}\, \mathrm{Tr}\,\Big( \bar{B}\, \gamma_5 \gamma^\mu \{\partial_\mu \Phi,B\} \Big)
\nonumber\\
&-& \frac{C}{2f}\, \mathrm{tr}\,\Big(\bar{\Delta}_\mu \cdot (\partial_\nu \Phi)\,
\big[g^{\mu\nu}-\frac{1}{2}\,Z\,\gamma^\mu \,\gamma^\nu\big]\, B + \mathrm{h.c.} \Big)
\nonumber\\
&-& \frac{H}{2f} \mathrm{tr} \,\Big(\big[ \bar{\Delta}^\mu \cdot  \gamma_5\,\gamma_\nu\,
\Delta_\mu \big]\,(\partial^\nu \Phi)\,\Big)\,,
\label{chiral-FD}
\end{eqnarray}
where we apply the notations of \cite{Lutz:Kolomeitsev:2002}. We use $f = 92.4$ MeV  in this work.
The values of the coupling constants $F,D,C$ and $H$ may be correlated by a large-$N_c$ operator
analysis \cite{Dashen,Jenkins,Jenkins:Manohar}. At leading order the coupling constants can be expressed in
terms of $F$ and $D$ only. We employ the values for $F$ and $D$ as suggested in \cite{Okun,Lutz:Kolomeitsev:2002}.
All together we use
\begin{eqnarray}
&& F = 0.45 \,, \qquad D= 0.80 \,, \qquad
 H= 9\,F-3\,D \,,\qquad C=2\,D \,,
\label{large-Nc}
\end{eqnarray}
in this work. We take the parameter $Z=0.72$ from a detailed coupled-channel study of meson-baryon
scattering that was based on the chiral SU(3) Lagrangian \cite{Lutz:Kolomeitsev:2002}. The latter parameter
is an observable quantity within the chiral SU(3) approach: it contributes at
order $Q^2$ to the meson-baryon scattering amplitudes but cannot be absorbed into the available
$Q^2$ counter terms \cite{Lutz:Kolomeitsev:2002}.

\section{Chiral loop expansion at the one-loop order}

We briefly recall the results of \cite{Semke:Lutz:2005}. A summation approach was defined by
performing a chiral loop expansion rather than a strict
chiral expansion: for a given truncation of the relativistic chiral Lagrangian
we take the loop expansion that is defined in  terms of the approximated Lagrangian seriously.
The number of loops we would consider is correlated with the chiral order to which the Lagrangian
is constructed. A renormalization based on the Passarino-Veltman reduction was devised that
installs the correct minimal chiral power of a
given loop function. The residual dependence on the renormalization scales is used to monitor the convergence
properties of the expansion and therewith estimate the error encountered at a given truncation.

The loop contribution to the baryon octet and decuplet masses read:
\begin{eqnarray}
&&\Delta M^{\rm loop}_{B \in [8]} = \sum_{Q\in [8], R\in [8]}
\left(\frac{G_{QR}^{(B)}}{2\,f} \right)^2  \Bigg\{
\frac{M_R^2-M_B^2}{2\,M_B}\, \bar I_Q
\nonumber\\
&& \quad - \frac{(M_B+M_R)^2}{E_R+M_R}\, p^2_{QR}\,
\Big(\bar I_{QR} + \frac{\bar I_Q}{M_R^2-m_Q^2}\,\Big)\Bigg\}
\label{octet}\\
&& \qquad \quad \;\,\,+\sum_{Q\in [8], R\in [10]}
\left(\frac{G_{QR}^{(B)}}{2\,f} \right)^2 \, \Bigg\{
\Bigg( \frac{(M_R-M_B)\,(M_R+M_B)^3+m_Q^4}{12\,M_B\,M^2_R}\,
\nonumber\\
&& \quad
- \frac{(Z\,(Z+2)-5)\,M_B^2+2\,(2\,Z\,(Z-1)-3)\,M_R\,M_B+2\,M_R^2}{12\,M_B\,M_R^2}\,m_Q^2\Bigg)\,
\bar I_Q
\nonumber\\
&& \quad  - \frac{2}{3}\,\frac{M_B^2}{M_R^2}\,\big(E_R+M_R\big)\,p_{QR}^{\,2}\,
\Big(\bar I_{QR} + \frac{\bar I_Q}{M_R^2-m_Q^2}\Big) \Bigg\}
\,, \nonumber
\end{eqnarray}
and
\begin{eqnarray}
&&\Delta M^{\rm loop}_{B\in [10]} = \sum_{Q\in [8], R\in [8]}
\left(\frac{G_{QR}^{(B)}}{2\,f} \right)^2  \Bigg\{
\Bigg( \frac{(M_R-M_B)\,(M_R+M_B)^3+m_Q^4}{24\,M^3_B}\,
\nonumber\\
&& \quad
- \frac{3\,M_B^2+2\,M_R\,M_B+2\,M_R^2}{24\,M^3_B}\,m_Q^2\Bigg)\,
\bar I_Q
\nonumber\\
&& \quad -\frac{1}{3}\,\big( E_R +M_R\big)\,p_{QR}^{\,2}\,
\Big(\bar I_{QR}+ \frac{\bar I_Q}{M_R^2-m_Q^2}\Big) \Bigg\}
\label{decuplet}
\\
&& \qquad \quad \;\,\,+\sum_{Q\in [8], R\in [10]}
\left(\frac{G_{QR}^{(B)}}{2\,f} \right)^2 \, \Bigg\{ \Bigg( \frac{(M_B+M_R)^2\,m_Q^4}{36\,M_B^3\,M_R^2}
\nonumber\\
&&\quad
+\frac{3\,M_B^4-2\,M^3_B\,M_R+3\,M_B^2\,M_R^2-2\,M_R^4}{36\,M_B^3\,M_R^2}
\,m_Q^2
\nonumber\\
&&\quad +\frac{M_R^4+M_B^4+12\,M_R^2\,M_B^2-2\,M_R\,M_B\,(M_B^2+M_R^2)}{36\,M^3_B\,M^2_R}\,
(M^2_R-M^2_B)
\Bigg)\,\bar I_Q
\nonumber\\
&& \quad  -\frac{(M_B+M_R)^2}{9\,M_R^2}\,\frac{2\,E_R\,(E_R-M_R)+5\,M_R^2}{E_R+M_R}\,
p_{QR}^{\,2}\,\Big(\bar I_{QR}(M_B^2)+ \frac{\bar I_Q}{M_R^2-m_Q^2}\Big)  \Bigg\}\,,
\nonumber
\end{eqnarray}
where
\begin{eqnarray}
&& \bar I_Q =\frac{m_Q^2}{(4\,\pi)^2}\,
\ln \left( \frac{m_Q^2}{\mu_{\,UV}^2}\right)\,,
\nonumber\\
&& \bar I_{Q R}=\frac{1}{16\,\pi^2}
\left\{ \frac{2\,\pi\,\mu_{IR}}{M_R}+\left(\frac{1}{2}\,\frac{m_Q^2+M_R^2}{m_Q^2-M_R^2}
-\frac{m_Q^2-M_R^2}{2\,M_B^2}
\right)
\,\ln \left( \frac{m_Q^2}{M_R^2}\right)
\right.
\nonumber\\
&& \;\quad \;\,+\left.
\frac{p_{Q R}}{M_B}\,
\left( \ln \left(1-\frac{M_B^2-2\,p_{Q R}\,M_B}{m_Q^2+M_R^2} \right)
-\ln \left(1-\frac{M_B^2+2\,p_{Q R}\,M_B}{m_Q^2+M_R^2} \right)\right)
\right\}\;,
\nonumber\\
&& p_{Q R}^2 =
\frac{M_B^2}{4}-\frac{M_R^2+m_Q^2}{2}+\frac{(M_R^2-m_Q^2)^2}{4\,M_B^2} \,,\qquad
E_R^2=M_R^2+p_{QR}^2 \,.
\end{eqnarray}
The sums in (\ref{octet}, \ref{decuplet}) extend over the intermediate Goldstone bosons ($Q\in[8]$) baryon
octet ($R\in [8]$) and decuplet  states ($R\in[10]$). The various coupling constants $G_{QR}^{(B)}$  are
determined by the parameters $F,D,C,H$. They are listed in \cite{Semke:Lutz:2005}. We emphasize that
(\ref{octet}, \ref{decuplet}) depend on the physical meson and baryon masses
$m_Q$ and $M_{R}$. This defines a self consistent summation since the masses of the intermediate baryon states
in (\ref{octet}, \ref{decuplet}) should match the total masses. At order $Q^3 $ the latter are the sum of the
tree-level contributions linear in the parameter $b_{0},b_D, b_F$ or $d_0, d_D$ and the loop contribution
(\ref{octet}, \ref{decuplet}).

The mesonic tadpole $\bar I_Q$  enjoys a logarithmic
dependence on the ultraviolet renormalization
scale $\mu_{UV}$ and the one-loop master function  $\bar I_{QR}$ a linear dependence on the infrared
renormalization scale $\mu_{IR} $. By construction the results (\ref{octet}, \ref{decuplet}) are
necessarily consistent with all chiral Ward identities as discussed in \cite{Semke:Lutz:2005}.
As emphasized by the authors the parameters $b_0, b_D$, $b_F$ and
$d_0, d_D$ are strongly dependent on the infrared scale $\mu_{IR}$. Applying a further chiral expansion
to (\ref{octet}, \ref{decuplet}) it was demonstrated that the physical masses are scale independent as they
should be. However, the convergence properties reflect the choice of the renormalization scales.
Good convergence properties can only be expected for natural values thereof. For the infrared scale
$\mu_{IR} \sim Q$ a natural window  350 MeV $< \mu_{IR} < $ 550 MeV was suggested in \cite{Semke:Lutz:2005}.
Keeping the partial summation as defined by (\ref{octet}, \ref{decuplet})
a residual dependence on the renormalization scales remains. This is
analogous to the residual cutoff dependence of the scheme of Donoghue and Holstein \cite{Donoghue:Holstein:1998}.
As long as such dependencies are small and decreasing as higher order terms are included they do not pose
a problem, rather, they offer a convenient way to estimate the error encountered at a given truncation.

\section{Quark-mass dependence of the baryon masses}

We discuss the implications of (\ref{octet}, \ref{decuplet}) for chiral extrapolations of
lattice simulations of the baryon masses \cite{MILC:2001,MILC:2004}.
Since we are not aiming at a chiral extrapolation
of the lattice simulations down to physical quark masses we adjust part of the parameters to
empirical data directly. The goal of the present study is a qualitative understanding
of the quark-mass dependence of the baryon masses.

For given values of the infrared and ultraviolet renormalization scales
the parameters $b_D, b_F$ and $d_D$ are fitted to the mass differences of the
octet states and decuplet states. The absolute mass scale of the octet and decuplet states can be reproduced
by appropriate values of the bare baryon masses. This procedure leaves undetermined the two parameters
$b_0$ and $d_0$. Good representations of
the physical baryon masses can be obtained for a wide range of the latter parameters. The parameter
$b_0$ may be used to dial a given pion-nucleon sigma term at physical pion masses. Similarly the
unknown parameter $d_0$ may be determined as to reproduce a given pion-delta sigma term.

\begin{table}[t]
\begin{center}
\begin{tabular}{|c|c|c|c|}\hline
 & $\mu_{IR}=350$\, MeV &  $ \mu_{IR}=450$\, MeV &  $ \mu_{IR}=550$\,
MeV \\ \hline \hline
$b_0\; \mathrm{[GeV^{-1}]}$ & $-0.89$ & $-0.63$ & $-0.38$\\
$b_D\; \mathrm{[GeV^{-1}]}$ & $+0.29$ & $+0.19$ & $+0.10$\\
$b_F\; \mathrm{[GeV^{-1}]}$ & $-0.34$ & $-0.25$ & $-0.15$\\
$d_0\; \mathrm{[GeV^{-1}]}$ & $-0.22$ & $-0.15$ & $-0.08$\\
$d_D\; \mathrm{[GeV^{-1}]}$ & $-0.35$ & $-0.30$ & $-0.24$\\
\hline \hline
$M_N$ [MeV] & $\phantom{1}750+\phantom{1}310-\phantom{1}121  $& $
\phantom{1}813+\phantom{1}232-\phantom{1}105 $ &
$ \phantom{1}875+\phantom{1}153-\phantom{11}89 $\\
 & $=\phantom{1}939  $& $ =\phantom{1}939 $ & $ =\phantom{1}939 $\\
$M_\Lambda$ [MeV] & $\phantom{1}750+\phantom{1}536-\phantom{1}150  $&$
\phantom{1}813+\phantom{1}398-\phantom{111}79 $
&$ \phantom{1}875+\phantom{1}260-\phantom{111}9 $\\
 & $=1136  $&$ =1131 $ &$ =1126 $\\
$M_\Sigma$ [MeV] & $\phantom{1}750+\phantom{1}882-\phantom{1}425 $ &
$\phantom{1}813+\phantom{1}630-\phantom{1}239 $
& $\phantom{1}875+\phantom{1}379-\phantom{11}54 $\\
 & $=1207 $ & $=1203 $ & $=1200 $\\
$M_\Xi$ [MeV] & $\phantom{1}750+\phantom{1}934-\phantom{1}361 $&
$\phantom{1}813+\phantom{1}680-\phantom{11}171 $
& $\phantom{1}875+\phantom{1}427+\phantom{11}19 $\\
& $=1323 $& $=1321 $ & $=1320 $\\
\hline \hline
$M_\Delta$ [MeV] & $1082+\phantom{1}241-\phantom{11}91  $&
$1108+\phantom{1}164-\phantom{11}39$
& $1133 +\phantom{11}87+\phantom{11}12$\\
 & $=1232  $& $ =1232 $ & $ =1232 $\\
$M_\Sigma$ [MeV] & $1082+\phantom{1}347-\phantom{11}49  $&$
1108+\phantom{1}253+\phantom{11}17 $
&$1133+\phantom{1}160+\phantom{11}83 $\\
 & $=1380  $&$ =1378 $ &$ =1376 $\\
$M_\Xi$ [MeV] & $1082+\phantom{1}453-\phantom{111}4 $ &
$1108+\phantom{1}343+\phantom{11}78 $
& $1133+\phantom{1}233+\phantom{1}160 $\\
 & $=1530 $ & $=1528 $ & $=1526 $\\
$M_\Omega$ [MeV] & $1082+\phantom{1}558+\phantom{11}34 $&
$1108+\phantom{1}432+\phantom{1}134$
& $1133+\phantom{1}305+\phantom{1}235$\\
 & $=1674 $& $=1674 $ & $=1674 $\\
\hline \hline
$\sigma_{\pi N}$ [MeV] & 52.4 & 53.9 & 55.7\\
\hline
$\sigma_{K^- p}$ [MeV] & 384.0 & 380.1 & 386.3\\
\hline
$\sigma_{K^- n}$ [MeV] & 359.4 & 354.8 & 361.5\\ \hline
\end{tabular}
\caption{The parameters are fitted so that (\ref{octet}, \ref{decuplet})
reproduces the baryon masses at
physical pion masses as well as the SU(3) limit values $M_{[8]} \simeq
1575$ MeV and $M_{[10]}\simeq 1710$ MeV
at $m_{\pi}\simeq 690$ MeV. We use $\mu_{UV}=800$ MeV.
The masses are decomposed into their chiral moments. }
\label{tab:parameter}
\end{center}
\end{table}

In this work $b_0$ and $d_0$ are adjusted as to reproduce the baryon octet and decuplet masses
in the SU(3) limit at $m_{\pi}\simeq 690$ MeV. The MILC
simulations suggest the values $M_{[8]} \simeq 1575$ MeV and $M_{[10]}\simeq 1710$ MeV.
This procedure is biased to the extent that it assumes that the chiral one-loop results
will be applicable at such high quarks masses. On the other hand  as long as there are no continuum
limit results of the MILC collaboration available this is an economical way to minimize the
influence of lattice size effects. The latter are expected to be smaller at large quark masses.
The procedure may be justified in retrospect if it turns out that the extrapolation recovers the
behavior predicted by the lattice simulation.

The parameters used in this work are collected in Tab. \ref{tab:parameter} together with the implied
masses of the baryon octet and decuplet states. A fair representation of the physical baryon masses
is obtained. For simplicity the values quoted in Tab. \ref{tab:parameter} correspond to a computation
where the intermediate baryon masses are put to their empirical values. Since the resulting masses are
very close to the physical masses this is well justified. As discussed in detail in \cite{Semke:Lutz:2005} the
parameters $b_{0,D,F}$ and $d_{0,D}$ show a strong
dependence on  the infrared scale $\mu_{IR}$. In contrast the physical baryon masses suffer from a weak
dependence only. For natural values of the infrared scale the chiral expansion appears well converging
as indicated by the decomposition of the baryon masses into their moments.

We emphasize that the values of the pion-nucleon sigma term shown in Tab. \ref{tab:parameter}
are an immediate consequence of the MILC simulation of the SU(3) symmetric point defined by
$m_\pi\simeq 690$ MeV
as discussed above. No further results of the MILC simulation are used. Taking the residual scale
dependence of the pion-nucleon sigma term as a naive error estimate we obtain
$\sigma_{\pi N} = 54 \pm 2$ MeV. The latter value is somewhat larger than the canonical value
of Gasser, Leutwyler and Sainio $\sigma_{\pi N} \simeq 45$ MeV \cite{Gasser:Leutwyler:Sainio:1991}.
Our value is, however, quite compatible with a recent analyses of  Frink and
Meissner who suggest $\sigma_{\pi N} \simeq 52$ MeV \cite{Frink:Meissner:2005}.
Also  Pascalutsa and Vanderhagen \cite{Pascalutsa:Vanderhagen:2005} arrive
at a somewhat larger value $\sigma_{\pi N} \simeq 57$ MeV based on a chiral SU(2) analysis of the
MILC simulation points.

\begin{figure}[t]
\begin{center}
\includegraphics[width=15cm,clip=true]{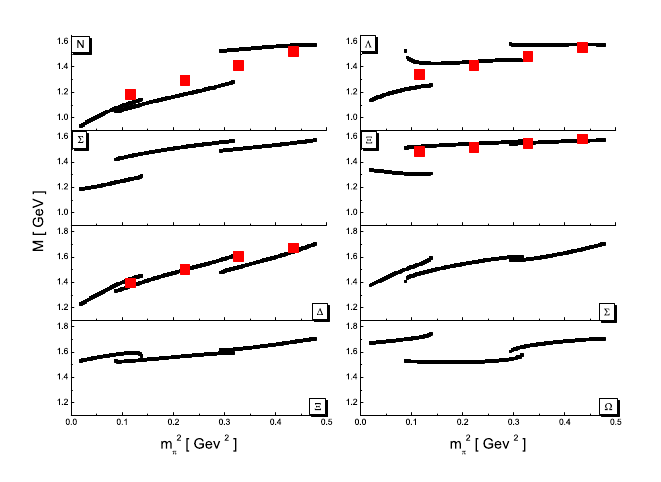}
\end{center}
\caption{The pion mass dependence of baryon octet and decuplet masses predicted by the chiral
loop expansion taking the parameters of Tab. \ref{tab:parameter}. The  lines represent the
masses for the infrared scale put at $\mu_{IR}=450$ MeV. The solid squares
are the simulation points of the MILC collaboration. }
\label{fig:1}
\end{figure}

We turn to the pion-mass dependence of the baryon octet and decuplet masses. Our results are collected in
Fig. \ref{fig:1} based on the parameters of Tab. \ref{tab:parameter}. We show results for the particular choice
$\mu_{IR}= 450$ MeV. It should be emphasized that the baryon masses are a solution of a set of coupled and
non-linear equations in the present scheme. This is a consequence of self consistency imposed on the
partial summation approach. The latter is a crucial requirement since the loop functions depend sensitively on the
precise values of the baryon masses. As a consequence of the non-linearity  for a given parameter set there
is no guarantee for a unique solution to exist, nor that solutions found are continuous in the quark masses.
Indeed as illustrated by Fig. 1 the pion-mass dependence predicted by the chiral loop expansion
is quite non-trivial exhibiting various discontinuities. Upon inspecting the energy dependence of the
baryon self energies this is readily understood: the baryon self energies, as implied by
(\ref{octet}, \ref{decuplet}), allow in certain cases for multiple solutions of the mass equation,
\begin{eqnarray}
\pslash=\chiral{M}+\Sigma_B(\pslash) \,,
\label{mass-equation}
\end{eqnarray}
even though the self consistency requirement singles out a particular solution at $\pslash=M_B$.
Such a scenario appears inconsistent at first: the meson baryon loop
functions have to be evaluated with all states provided by (\ref{mass-equation}). In contrast the computation
of the meson-baryon loop functions (\ref{octet}, \ref{decuplet}) takes the existence of uniquely defined
baryon states for granted. In addition
it is assumed that the mass poles of the baryon propagators have residuum one. Thus our computation
considers implicitly additional counter terms of the form:
\begin{eqnarray}
\Sigma_{B}^{wave-function}(\pslash) = \zeta_B\,(\pslash-M_B )\,,
\label{wave-function}
\end{eqnarray}
where $M_B$ denotes the baryon masses that solve the self consistent system. If the term (\ref{wave-function})
is added to the self energies it does not affect the solutions of the self consistent system. However, it does affect the energy dependence
of the final baryon self energy. By adjusting the parameters $\zeta_B$, it is possible to arrive at
uniquely defined  baryon states.

Within the scheme described above it is intuitive that for a given
set of parameters multiple solutions of the self consistent system
may exist. A priori it is not evident which of the possible
solutions one should select. Our strategy is to continuously
evolve the solution established at physical pion masses to larger
pion masses. Similarly we evolve down to smaller pion masses the
solution that reproduces the lattice simulations at the SU(3)
symmetric point defined by $m_\pi\simeq $ 690 MeV. Both solutions
reach end points beyond which we do not find any continuation.
Mathematically the endpoints are characterized by the divergence of the
sigma term. In the region of intermediate pion masses we,
however, find another solution, suggesting a discontinuous
dependence of the baryon masses on the quark masses of QCD. One
may expect that the physical solutions are those of lowest mass
always. The three branches discussed above are shown in Fig. 1 and
confronted with the simulation points of the MILC collaboration
\cite{MILC:2001,MILC:2004}.

It is striking to see that we reproduce the 'mysterious' pion-mass
dependence of the $\Xi$ mass, i.e. the quite flat behavior which does not seem to smoothly approach the
physical mass. Given the present uncertainties from finite lattice spacing, the staggered approximation and
the theoretical uncertainties implied by higher order contributions, we would argue
that we arrive at a fair representation of the lattice simulation points for all baryons with some reservation
concerning the nucleon. We note that the parameter set
used before in \cite{Semke:Lutz:2005} leads to results qualitative similar to those shown in Fig. 1 for
the baryon octet masses. While the description of the nucleon is improved showing no
discontinuity at large pion masses, the sizeable jump of the $\Xi$ mass at small pion masses remains.
However, the MILC simulation points of the isobar mass disfavor the
latter choice. An overestimate of the lattice simulations of the isobar mass at larger quark masses
would be the consequence. We checked that reasonable variations of the parameter set
does not change the results qualitatively. We could not find
parameters that lead to a smooth chiral extrapolation.

Incorporating the many $Q^4$ counter terms offered by the chiral Lagrangian it is
reasonable to expect that the latter will further improve the picture.
However, as long as there is no detailed analysis available
that performs the continuum limit and estimates the uncertainty from the staggered fermion approximation
there is not much point considering the $Q^4$ counter terms.

\section{Summary}

We evaluated the baryon octet and decuplet self energies at the one-loop level applying
the relativistic chiral Lagrangian. Adjusting the $Q^2$ counter terms an excellent representation
of the physical baryon octet and decuplet masses is obtained. The size of the pion-nucleon sigma term
was estimated by including in the fit the baryon octet and decuplet masses
at the SU(3) symmetric point defined by $m_\pi\simeq 690$ MeV. For the latter  the MILC collaboration
suggests the values $M_{[8]} \simeq 1575$ MeV and $M_{[10]}\simeq 1710$ MeV. As a result we predict
$\sigma_{\pi N} \simeq 54$ MeV. The kaon-nucleon sigma terms
are $\sigma_{K^-p} \simeq 380$ MeV and $\sigma_{K^-n}\simeq 355 $ MeV.

Given this scenario the pion-mass dependence of the baryon octet
and decuplet masses were evaluated. The latter are a solution of a set of coupled and non-linear
algebraic equations. This is a direct consequence of self consistency imposed on the partial summation, i.e.
the masses used in the loop functions are identical to those obtained from the baryon self energies.
As a striking consequence we predict a discontinuous dependence of the baryon masses on the pion mass.
Typically the baryon masses jump at pion masses as low as 300 MeV. Most spectacular is the behavior of the
$\Xi$ mass. At small pion masses it decreases with increasing pion masses. At a
critical pion  mass of about 300-400 MeV it jumps
up to a value amazingly close to the prediction of the MILC collaboration. For all baryon masses for which
the MILC collaboration published simulation points our results are reasonably close to the lattice
estimates, given the present uncertainties from finite lattice spacing, the staggered approximation and
the theoretical uncertainties implied by higher order contributions to the baryon self energies.

It is interesting to speculate on the physics behind such an unexpected
quark-mass dependence of the baryon masses. A discontinuous behavior
may signal a new type of phase transition - or perhaps only some
so-far unknown instability of QCD for certain parameter choices. One
may argue that in the intermediate quark-mass region a ghost state
appears causing an instability. At present we can not exclude the
possibility that our results merely indicate that the chiral extrapolation
stops making sense at quite small quark masses. On the other hand
further chiral correction terms may connect the various branches in a
smooth manner providing a quite non-liner quark mass dependence of the
baryon masses.

The answer to these questions is outside the scope of the present work.
However, we do not see any argument that this can not be and therefore
take our results as a quest for further detailed studies.

\section*{Acknowledgments}

M.F.M.L. acknowledges fruitful discussions with B. Friman and Ch.
Redlich.

\end{document}